\documentclass[aps,prb,twocolumn,groupedaddress,showpacs]{revtex4-1}

\usepackage{graphicx}
\begin{document}
\title{Mechanisms of Ignition by Transient Energy Deposition:
Regimes of Combustion Waves Propagation}
\author{A.D. Kiverin$^1$}
\email{alexeykiverin@gmail.com}
\author{D.R. Kassoy$^2$}
\author{M.F. Ivanov$^1$}
\author{M.A. Liberman $^{3,4}$}
\email{misha.liberman@gmail.com}

\medskip
\affiliation{$^1$ Joint Institute for High Temperatures,
Russian Academy of Science, Izhorskaya 13,
Bd. 2 Moscow 125412, Russia \\
$^2$ Department of Mechanical Engineering, Box 427,
University of Colorado, Boulder, Colorado 80309, USA\\
$^3$ Moscow Institute of Physics and Technology, Dolgoprudnyi, 141700, Russia\\
$^4$ Nordita, KTH Royal Institute of Technology and Stockholm University, Roslagstullsbacken 23, 10691 Stockholm, Sweden}

\date{\today}
\begin{abstract}
Regimes of chemical reaction wave propagating
in reactive gaseous mixtures, whose chemistry
is governed by chain-branching kinetics, are
studied depending on the characteristics of a
transient thermal energy deposition localized
in a finite volume of reactive gas. Different
regimes of the reaction wave propagation are
initiated depending on the amount of deposited
thermal energy, power of the source and the
size of the hot spot. The main parameters
which define regimes of the combustion waves
facilitated by the transient deposition of
thermal energy are: acoustic timescale,
duration of the energy deposition,
ignition time scale and size of the hot spot.
The interplay between these parameters
specifies the role of gasdynamical processes,
the formation and steepness of the temperature
gradient and speed of the spontaneous wave.
The obtained results show how ignition of
one or another regime of combustion wave
depends on the value of energy, rate of
the energy deposition and size of the hot
spot, which is important for the practical
use and for risk assessment.
\end{abstract}


\maketitle

\section{Introduction}

The initiation or ignition of a chemical reaction
is one of the most important and fundamental
problems in combustion physics. One needs to
know how combustion starts and how the
transient energy deposition influences the
regime of the reaction wave which propagates
out from a finite volume of reactive gas
where a transient thermal energy were
deposited - the hot spot. What type of
combustion wave is formed depending on:
the amount of energy actually added to a
finite volume of reactive gas on a specific
time scale, the power deposition, and the
ignition conditions, e.g. size of the hot
spot, initial pressure, etc.? Long ago
Oppenheim and Soloukhin \cite{OppenhSol1}
recognized the importance of these concepts
with the remark "Gasdynamics of Explosions is best
defined as the science dealing with the
interrelationship between energy transfer
occurring at a high rate in a compressible
medium and the concomitant motion set up in this medium".
The community of scholars has sought to address this
perspective for many years represented by a vast
combustion science literature too extensive to enumerate here.

Transient thermal energy deposition into a reactive
gas provides a source for ignition of either
deflagration or detonation. Sufficiently fast
and large energy addition can facilitate direct
initiation of detonation. However the particular
mechanism of the direct initiation of detonation
can be different. Detonation can be initiated by
a strong shock (strong explosion) or it can arise
as a result of the formation of an appropriate
temperature gradient through the Zeldovich' gradient mechanism
\cite{Zeld19802}.
In most practical cases ignition arises from a small volume
of combustible mixture which is locally heated by energy
input by means of an electric spark, hot wire, focused
laser light and other related external sources. Such a
transient energy addition process can generate a wide range of
gas expansion processes depending on the amount and the rate of
energy actually added and may result in the formation of the
initial non-uniform distribution of temperature
(see for example, Kassoy \cite{Kas3,Kas4}
for the diverse range of fluid responses to localized,
spatially distributed, thermal power addition into an inert gas).
An example of an initial nonuniform distribution of
temperature arises from the energy deposition of a
spark-plug in an engine combustor \cite{Echekki5}.
In the general case it can be non-uniform distributions
of temperature, pressure and/or concentration of
reactants which determine further evolution of
the reaction wave depending on the mixture
reactivity and initial pressure. An example of
concentration non-uniformity is a hydrogen gas
leakage and its nonuniform distribution by
convective mixing in a room. In all cases, a
reaction wave arises from the induction
time non-uniformity via the thermal explosion.

The ignition problem is important for
improving combustion safety and risk assessments
of processes where hydrocarbons are oxidized at
different initial conditions (concentration,
temperature and pressure). How can we minimize "accidental"
explosions in mines, chemical industry and nuclear
power plants? An important problem of "hydrogen safety"
is connected with leakage of hydrogen gas, subsequent
mixing with air and the mixture explosion due to
local heat release. It is worth noting that the
problem in question is also of great interest for
hydrogen storage, transportation and utilization
and for the design and operation of perspective
pulse detonation engines and homogeneous
charge compression ignition (HCCI) engines.

For the first time possible regimes of
chemical reaction wave ignited by the initial
non-uniform distribution of temperature have
been studied by Ya. B. Zeldovich using a
one-step chemical reaction model \cite{Zeld19802}.
The basic idea of the Zeldovich's concept was
that a spontaneous reaction wave can propagate
through a reactive material along a spatial
gradient of temperature $\bf \nabla T(x)$,
with the velocity
\begin{eqnarray}
U_{sp} = {\left| {\left( {d{\tau _{ind}}/dx} \right)}
\right|^{ - 1}} = \left| {{{\left( {\partial {\tau _{ind}}/\partial T}
\right)}^{ - 1}}{{\left( {\partial T/\partial x} \right)}^{ - 1}}} \right|
 \label{eq1}
\end{eqnarray}
where ${\tau _{ind}}(T(x))$ is the induction time.
The value of ${U_{sp}}$ depends only on the steepness
of the temperature gradient. Then the regime of the
formed combustion wave depends on the value of
spontaneous wave velocity compared to the sound speed.

Recently regimes of chemical reaction wave
propagation initiated by initial temperature
non-uniformity in gaseous mixtures, whose chemistry
is governed by chain-branching kinetics, were studied
using a multispecies transport and detailed chemical
model \cite{LKI6}. Possible regimes of the reaction wave
propagation were identified for the stoichiometric
hydrogen/oxygen and hydrogen/air mixtures in a wide
range of initial pressures and temperature level
depending on the initial gradient steepness.

The question that still remained unanswered is
how the temperature gradient in the Zeldovich's
concept of the spontaneous reaction wave \cite{Zeld19802}
arises. Kassoy and co-authors \cite{Kas7,Kas8,Kas9,Kas10,Kas11,Kas12,Kas13}
use a one step chemical
reaction model to study how a temperature distribution
adjacent to a planar boundary, generated either by
direct deposition of transient, spatially distributed
thermal power into a defined volume of reactive gas, or
by conduction through the boundary into the gas, leads
to planar detonation initiation. The authors seek to
understand the magnitude of the energy addition deposited
on specific time and length scales required to produce
conditions that will lead to detonation initiation.
The important difference between a one-step and detailed
chemical models is seen, for example, from the
result of \cite{Sloane14}, where it was shown the ignition
energy for methane-air computed using a one-step
model differs by two orders of magnitude from
the experimentally measured value.

Liberman et. al  \cite{LKI6} take a different approach to
show that steepness of an imposed temperature gradient
(the length scale   at fixed temperature difference)
required for initiating combustion regimes, in particular
a detonation, in gaseous mixtures with chain-branching
kinetics, may differ up to several orders of magnitude
from that obtained using a one step global chemical
reaction model. The energy of ignition for hydrogen-oxygen
gaseous mixture has been calculated in \cite{Warnatz15} using a
detailed reaction mechanism and a multispecies
transport model, but difference in ignition of different
combustion regimes remained unanswered.

The purpose of the present paper is to study
how the initial temperature gradient is formed
depending on the rate and amount of energy addition,
on the size of hot spot and on the gasdynamic processes
emerging in the region of transient energy deposition.
The problem of great practical importance is: how is
temperature non-uniformity initiating different
combustion regimes arise? How do the gasdynamic
processes caused by the energy addition influence
the formation of the temperature gradient and how it
depends on the rate and amount of energy input and
on the size of hot spot? Solution of this problem
answers the question of great practical importance:
what is the amount of energy and how it should be deposited
to ignite certain regimes of combustion wave.
The paper presents new results on classification of
the propagation regimes of chemical reaction wave
initiated by the transient energy deposition in
gaseous mixtures using high resolution numerical
simulations of reactive Navier-Stokes equations,
including a multispecies transport model, the
effects of viscosity, thermal conduction, molecular
diffusion and a detailed chemical reaction mechanism
for hydrogen-oxygen mixture which is the quintessential
example of chain branching reactions whose chemical kinetics
is well understood and whose detailed chemical kinetic models
are well known and relatively simple. Such a level of
modeling allows clear understanding of the feedback
between gasdynamics and chemistry, the principal
point when studying unsteady process of ignition,
not easily be captured using simplified gas-dynamical and chemical models.

High fidelity reliable numerical simulations of
the present study are performed to identify ignition
processes in homogeneous hydrogen-oxygen mixtures
caused by the localized energy deposition.
The distinct ignition regimes are identified
depending on the size of the hot spot, energy
amount, the duration of energy addition and
initial pressure. The obtained results open
perspectives for understanding of how to avoid
or on the contrary what are the conditions to
initiate different combustion regimes
(slow deflagration, fast deflagration, detonation).
The performed analysis reveals the conditions
when detonation is initiated as a result of
direct initiation by a strong shock wave, or
when it results from the formation of suitable
temperature gradient for the Zeldovich's
mechanism of detonation triggering.
The amount and the rate of energy addition
needed to form a proper gasdynamics nonuniformity
for triggering either detonation or deflagration waves are found.

\section{Problem setup}

We consider uniform initial conditions and a transient
external source of energy localized on the scale of the "hot spot"
$0 \le x \le L$, where energy ${Q_{ig}}$  is added
during the time $\Delta {t_Q}$.
Gasdynamics of the explosion initiated by the localized
energy deposition is characterized by the interrelationship
between the rate of energy transfer with the
time characterizing energy deposition $ \Delta {t_Q}$
in the hot spot of size $L$ and the characteristic
times of the problem. For the sake of simplicity we
assume that the rate of the energy addition is
constant in time, so that total energy deposition
into the hot spot is ${Q_{ig}} = W\Delta {t_Q}$,
where $W$ is the power of the external source
of energy, and  $\Delta {t_Q}$ is time of
the hot spot transient heating.
The characteristic acoustic time
${t_a} = L/a(T)$ defines the concomitant motion
setup in the gaseous mixture, where $a(T)$
is the speed of sound. If $\Delta {t_Q} \ll {t_a}$
local heat addition occurs as a nearly constant
volume process and the temperature elevation
within the hot spot is accompanied by a
concomitant pressure rise.  Subsequent expansion of
the hot spot driven by the large pressure gradient
between the hot spot and the ambient gas causes
compression and shock waves in the
surrounding ambient gas. When  $ \Delta {t_Q} \gg {t_a}$
the acoustic waves have enough time for pressure
equalization and the local heat
addition occurs at nearly constant pressure.

The local heat of gaseous mixture within
the hot spot leads to the heat being concentrated
in a small region from which heat propagates
into the surrounding gas according to well
known solution of the equation of thermal
conduction \cite{LL16}. For the one-dimensional
case the temperature distribution (for constant thermal diffusivity) is
\begin{eqnarray}
T(x,t) \propto \frac{1}
{{2\sqrt {\pi \chi t} }}\exp ( - {x^2}/4\chi t)
\label{eq2}
\end{eqnarray}
where $\chi  = \kappa (T)/\rho {C_P}$
is the coefficient of thermal diffusivity.
It follows from Eq.~(\ref{eq2}) that heat
propagates at the distance $x \sim \sqrt {4\chi t}$
and temperature in the surrounding gas increases
noticeably due to propagation of a thermal wave during the time
\begin{eqnarray}
{t_T} \sim {x^2}/\chi
\label{eq3}
\end{eqnarray}

At the very beginning the thermal
wave propagates with the velocity $dx/dt \sim \sqrt {\chi /t}$
overtaking the shock wave, which propagates
with a velocity approximately equal to the
sound speed in the heated gas.

However, very soon the shock wave overtakes the
thermal wave, so that characteristic thermal wave time
is effectively much longer than the acoustic time.
For example, ${t_T} \sim {x^2}/\chi  \approx 0.1$s
and ${t_a} \sim 2\mu$s  for $x \simeq 1$mm for
hydrogen-oxygen mixture at $P = 1$atm.

\begin{figure}
\vspace*{1mm} \centering
\includegraphics[width=9cm]{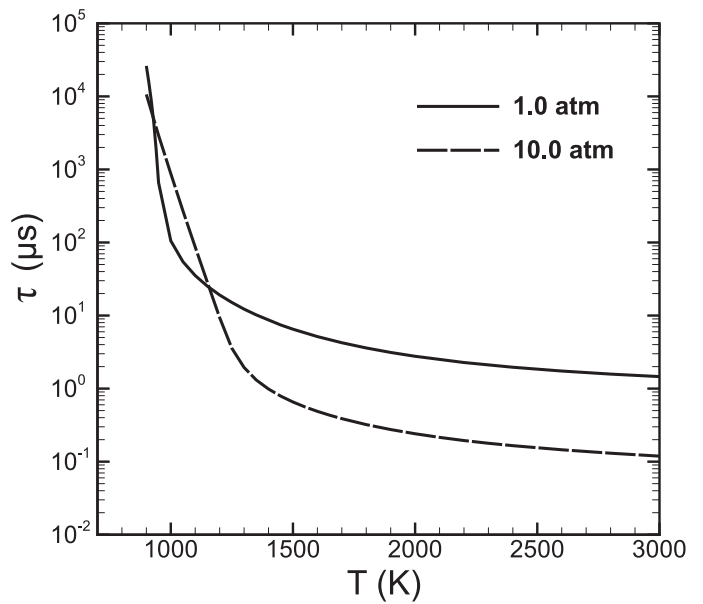}
\caption {\label{Fig1}
Induction time for hydrogen-oxygen
stoichiometric mixture at different
temperatures and pressures:  ${P_0} = 1atm$
 (solid line) and ${P_0} = 1atm$  (dashed line).}
\end{figure}

Whether a chemical reaction starts and which processes
define the regime of combustion wave depend on the
interrelationship between $\Delta {t_Q}$, ${t_a}$,
${t_T}$ and induction time ${t_{ind}}(T)$
at the existing temperature and pressure.
The induction time is the time scale for the
stage of endothermic chain initiation and
branching reactions (in the case of a global
one-step reaction this is the time-scale
for the maximum reaction rate).
Here it is suitable to use the scale of ignition
time $t_{ign}$ , characterizing the length of
induction phase after or
during the transient energy deposition ${t_{ind}}(T,P)$.
The induction time is measured experimentally
and determines local properties of the combustible
mixture depending on its thermodynamic state.
Dependence of the induction time on temperature
for a hydrogen-oxygen mixture at
initial pressures ${P_0} = 1$atm  and ${P_0} = 10$atm is
shown in Fig. \ref{Fig1}.

For the temperature range $T = (1100 - 1500){\rm{K}}$,
where the exothermic reaction starts, the induction
time is about ten microseconds. If the reaction has
started, further heating and energy deposition do not
matter and do not influence the formed combustion regime.
Thus, primary interest is focused on the regimes of
ignition with  $\Delta {t_Q} < {t_{ign}}$.
For a very short time of energy addition, much
less than the acoustic time, the mixture in the
hot spot can be heated to any temperature and the
ignition regime will be determined by the induction
time at that temperature and accompanying pressure.
In case of a more extended energy deposition, the
ignition regime will depend on the size of the
hot spot (and correspondingly on $t_{a}$) and
the relation between $\Delta {t_Q}$ and $t_{a}$
as mentioned earlier in reference to the features
of the hot spot expansion process.

For a one-dimensional formulation the hot
spot represents a region $0 \le x \le L$,
where energy is deposited with the rate  $W(t) = dQ(t)/dt$.
The specific internal energy of the mixture is changed as
$Q(t) = Q(t - \Delta t) + \Delta Q$
within the hot spot during time interval $\Delta t$.
Total energy at the end of the energy addition depends
on the gas-dynamic motion

\begin{eqnarray}
Q_{total} = \sum\limits_{t = 0}^{\Delta {t_Q}}
{\sum\limits_{x = 0}^L {\Delta Q \cdot \rho (x,t) \cdot \Delta x} }.
 \label{eq4}
\end{eqnarray}

Total energy transfer till the formation
of a 1-D stationary combustion wave far
from the hot spot (at $x \gg L$) can be viewed
as the ignition energy required for initiation of
the particular combustion regime.

The governing equations are the one-dimensional
time-dependent, multispecies reactive Navier-Stokes
equations including the effects of compressibility,
molecular diffusion, thermal conduction, viscosity and
chemical kinetics with subsequent chain branching,
production of radicals and energy release.
\begin{eqnarray}
\frac{{\partial \rho }}{{\partial t}} +
\frac{{\partial \left( {\rho u} \right)}}{{\partial x}} = 0,
 \label{eq5}
\end{eqnarray}
\begin{eqnarray}
\frac{{\partial {Y_i}}}{{\partial t}} +
u\frac{{\partial {Y_i}}}{{\partial x}} =
\frac{1}{\rho }\frac{\partial }{{\partial x}}
\left( {\rho {D_i}\frac{{\partial {Y_i}}}
{{\partial x}}} \right) + {\left( {\frac{{\partial {Y_i}}}
{{\partial t}}} \right)_{ch}},
 \label{eq6}
\end{eqnarray}
\begin{eqnarray}
\rho \left( {\frac{{\partial u}}{{\partial t}} +
u\frac{{\partial u}}{{\partial x}}} \right) =
- \frac{{\partial P}}{{\partial x}} +
\frac{{\partial {\sigma _{xx}}}}{{\partial x}},
 \label{eq7}
\end{eqnarray}
\begin{eqnarray}
\rho \left( {\frac{{\partial E}}{{\partial t}} +
u\frac{{\partial E}}{{\partial x}}} \right) =
 - \frac{{\partial \left( {Pu} \right)}}{{\partial x}}
 + \nonumber\\*
 + \frac{\partial }{{\partial x}}\left( {{\sigma _{xx}}u} \right)
 +  \frac{\partial }{{\partial x}}
  \left( {\kappa \left( T \right)\frac{{\partial T}}{{\partial x}}}
   \right) + \nonumber\\*
   + \sum\limits_k {\frac{{{h_k}}}{{{m_k}}}
   \left( {\frac{\partial }{{\partial x}}
   \left( {\rho D_k^{}\left( T \right)\frac{{\partial {Y_k}}}
   {{\partial x}}} \right)} \right)}  + W(t),
 \label{eq8}
\end{eqnarray}
\begin{eqnarray}
P = {R_B}T\,n = \left( {\sum\limits_i
{\frac{{{R_B}}}{{{m_i}}}{Y_i}} } \right)\rho T
= \rho T\sum\limits_i {{R_i}{Y_i}}
 \label{eq9},
\end{eqnarray}
\begin{eqnarray}
\varepsilon  = {c_v}T + \sum\limits_k {\frac{{{h_k}{\rho _k}}}{\rho }}
= {c_v}T + \sum\limits_k {{h_k}{Y_k}},
 \label{eq10}
\end{eqnarray}
\begin{eqnarray}
{\sigma _{xx}} = \frac{4}{3}\mu
\left( {\frac{{\partial u}}{{\partial x}}} \right)
 \label{eq11}
\end{eqnarray}

The initial conditions at $t = 0$  are constant
pressure and zero velocity of the unburned mixture.
At the left boundary at $x = 0$ the conditions are
for a solid reflecting wall, where $u(0,t) = 0$
and the initial temperature $T(t = 0) = {T_0}$.

Here we use the standard notations:
$P$, $\rho $ , $u$, are pressure, mass density, and flow velocity,
${Y_i} = {\rho _i}/\rho$  - the mass fractions of the species,
$E = \varepsilon  + {u^2}/2 $ - the total energy density,
$\varepsilon $ - the internal energy density,
${R_B}$ - is the universal gas constant,
${m_i}$- the molar mass of i-species,
${R_i} = {R_B}/{m_i}$, $n$ - the molar density,
$ \sigma _{ij}$- the viscous stress tensor,
${c_v} = \sum\limits_i {{c_{vi}}} {Y_i}$ - is the
constant volume specific heat,
$c_{vi}$- the constant volume specific heat of i-species,
$h_i$  - the enthalpy of formation of i-species,
$\kappa (T)$  and $\mu (T)$  are the coefficients of thermal
conductivity and viscosity, ${D_i}(T)$  - is the diffusion coefficients
of i-species,  ${\left( {\partial {Y_i}/\partial t} \right)_{ch}}$ - is
the variation of i-species concentration
(mass fraction) in chemical reactions.

The equations of state for the reactive mixture and
for the combustion products were taken with the
temperature dependence of the specific heats
and enthalpies of each species borrowed from
the JANAF tables and interpolated by the
fifth-order polynomials \cite {Warnatz17,Heywood18}.
The viscosity and thermal conductivity coefficients
of the mixture were calculated from the gas kinetic
theory using the Lennard-Jones potential \cite {Hirschfelder19}
Coefficient of the heat conduction ${\kappa} = {\mu}{C_P}/\Pr $ for the mixture as a whole is expressed via the viscosity $\mu$ and the Prandtl number, $Pr  = 0.75$.

The numerical method is based on splitting of
the Eulerian and Lagrangian stages, known as
coarse particle method (CPM) \cite {Belotserkovsky20}.
A detailed description
of the modified CPM optimal approximation scheme,
details of the equations and transport coefficients
and the reaction kinetics scheme together with
the reaction rates appear in \cite {IKL21,IKL22}.
The numerical method is thoroughly tested and
successfully used in many practical
applications \cite {IKL22,LIV23,LIPV24}.

The convergence of the solutions
is of paramount importance to verify
that the observed phenomena are sufficiently
well resolved, especially when CFD simulations
are used with a detailed chemical reaction model.
The convergence and resolution tests have
shown that the resolution of 50
computational grid cells over the width
of a laminar flame (for example,
with the grid cell size $\Delta  = 0.0064$mm
at  ${P_0} = 1$atm, when the width of a laminar
front is 0.24mm, and much smaller for higher pressure)
provides sufficiently good convergence and correctly
captures the details of the observed processes
 (see Appendix in Ref.~\cite {LKI6}).

\section{Combustion regimes: rapid energy deposition - microsecond time scale}

We consider combustion regimes initiated by energy
deposition at the initial pressure and
temperature ${P_0} = 1$atm  and  ${T_0} = 300$K  in a hydrogen-oxygen
stoichiometric mixture. First we consider
cases when the time scale of the energy
deposition is comparable to or shorter than
the acoustic time scale and less than the induction
time at the ignition temperatures.
The acoustic time is ${t_a} \approx 20 \mu$s for the
hot spot of size $L = 1$cm, and ${t_a} \approx 2 \mu$s
for $L = 1$mm  (${a_0}(T = 300K) = 539$m/s).
Rapid energy addition into the hot spot on
time scales much shorter then acoustic time
causes almost uniformly fast elevation of
pressure and temperature resulting in
the volume explosion.
\begin{figure}
\vspace*{1mm} \centering
\includegraphics[width=9cm]{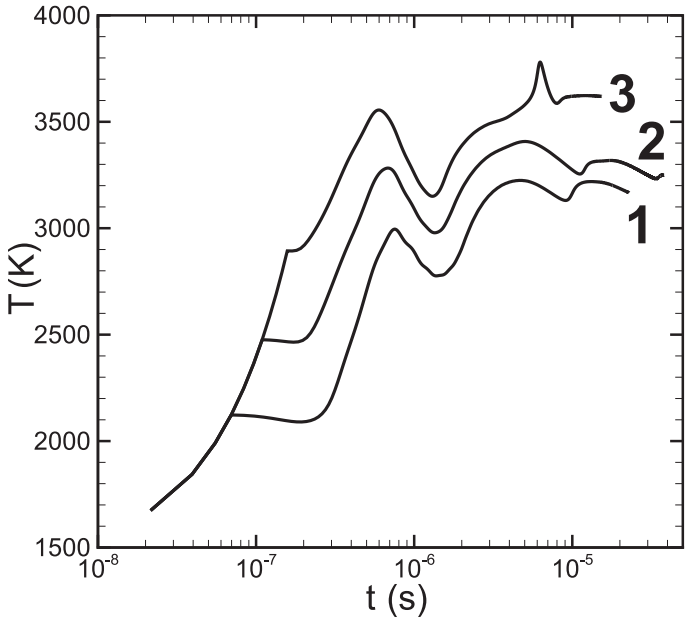}
\caption {\label{Fig2}
Temperature evolution in the center of the hot
spot for rapid submicrosecond energy deposition
$ \Delta {t_Q} = {\rm{(0}}{\rm{.1}} \div {\rm{0}}{\rm{.2)}} \mu {\rm{s}}$;
1 - deflagration ($Q = 1.9{kJ}/{m}^2$),
2 - deflagration ($Q = 2.4{kJ}/{m}^2$ ), 3 - detonation ($Q = 3.0{kJ}/{m}^2$ ).}
\end{figure}
Figure \ref{Fig2} shows the calculated temporal
evolution of temperature at the center
of the hot spot ($x = 0$) for different values
of the transmitted energy, when the energy
addition time is very short:
$ \Delta {t_Q} \ ~0.1\mu s < {t_{ign}} \ll {t_a}$.
Temperature and pressure of the mixture in
the hot spot depend on the energy transmitted
to the hot spot.
After the end of the energy
deposition process, the induction
period reaction starts. After about 10$\mu$s
the stationary combustion regimes are
established (see Fig.~\ref{Fig2}). It should be
noted that for each value of deposited
energy there is some definite temperature
and pressure at which the reaction starts
and the combustion regime is produced by the
volumetric explosion at these conditions.

\begin{figure}
\vspace*{1mm} \centering
\includegraphics[width=9cm]{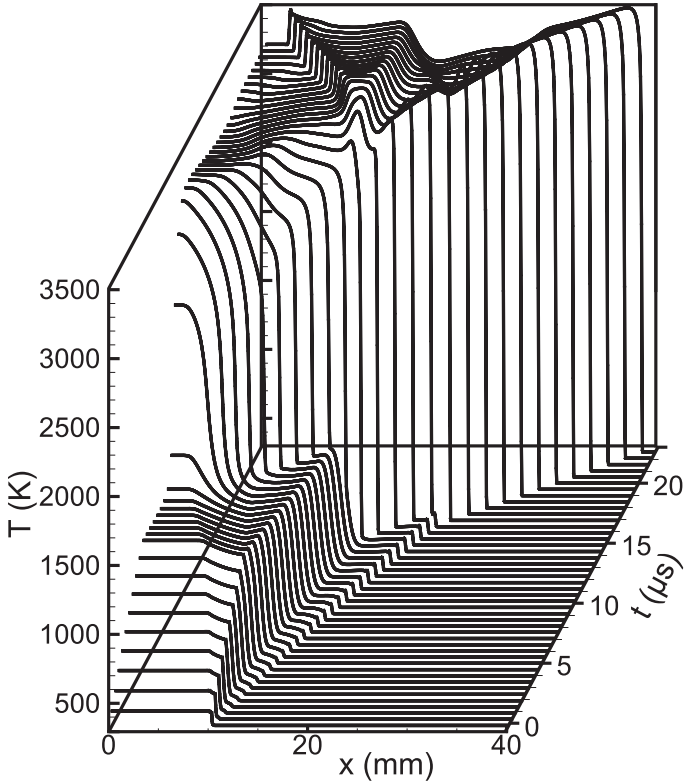}
\caption {\label{Fig3}
Evolution of temperature profiles illustrating detonation formation in the energy release region.
$L = 1$cm,  $ \Delta {t_Q} = 5\mu $s.
Profiles are presented for time instants with interval
$ \Delta {t_Q} = 0.5\mu $s.}
\end{figure}
\begin{figure}
\vspace*{1mm} \centering
\includegraphics[width=9cm]{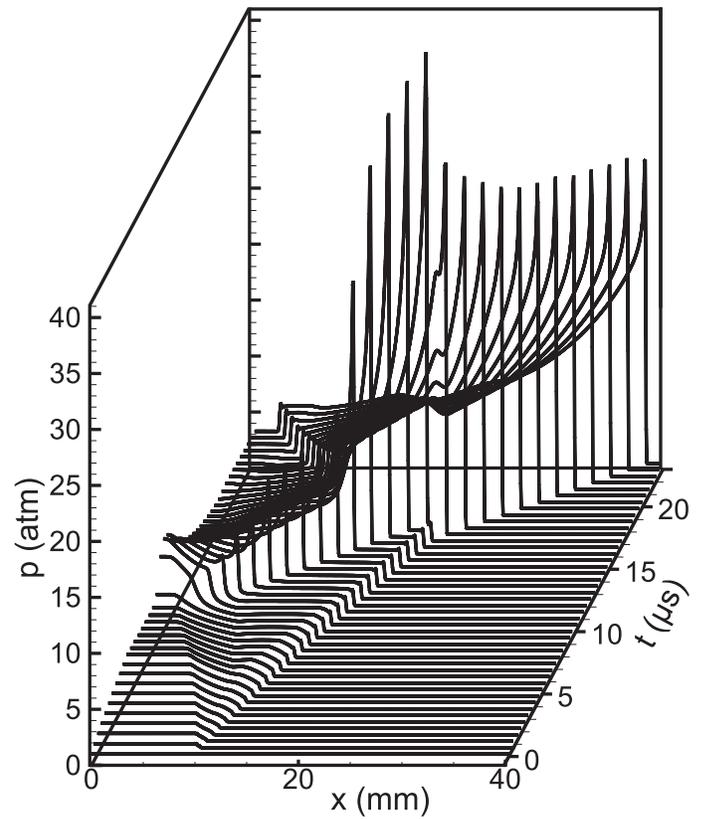}
\caption {\label{Fig4}
Evolution of pressure profiles illustrating detonation formation in the energy release region.
$L = 1$cm,  $ \Delta {t_Q} = 5\mu $s.
Profiles are presented for time instants with interval
$ \Delta {t_Q} = 0.5\mu $s.}
\end{figure}

For a less rapid process of energy deposition
($ \Delta {t_Q} = 5\mu s < {t_{ign}} < {t_a}$),
a large pressure jump is formed at the boundary of the hot spot.
If the power is large enough, subsequent events represent
the decay of the initial discontinuity consisting of
a compression wave propagating in the direction $x > 0$,
which steepens into the shock wave and the rarefaction
wave propagating to the left from the boundary of hot spot
(in the direction $x = 0$) with the velocity equal
to the local sound speed.
Such a scenario, which is similar to a
strong point explosion \cite {ZeldRaiz25,terVehn26}
results in the direct triggering of a detonation
wave if a concomitant shock wave at the right
boundary is strong enough.
A more interesting scenario emerges in the case
of a weaker shock and is shown in Fig.~\ref{Fig3} and Fig.~\ref{Fig4}.
These figures show the calculated transient evolution
of the temperature (Fig.~\ref{Fig3}) and pressure (Fig.~\ref{Fig4}) profiles inside the hot spot ($L = 1$cm) during the energy deposition.
The rarefaction wave propagating to the left creates
shallow temperature and pressure gradients on the
scale of about the size of the hot spot.
At an initial pressure of ${P_0} = 1$atm
the temperature gradient with the temperature
difference and length scale $L \sim 1$cm as in Figs.~\ref{Fig3},~\ref{Fig4}
cannot trigger detonation.
However, since the pressure of the heated mixture
increased during the heating up to $P \approx 4$atm,
this temperature gradient can produce a detonation
through the Zeldovich gradient mechanism.
Time evolution of the temperature
and pressure profiles shown in Figs.~\ref{Fig3},~\ref{Fig4} demonstrate
also the emergence of the spontaneous reaction wave
and its coupling with the pressure wave
leading to the detonation initiation
through the Zeldovich mechanism.

\begin{figure}
\vspace*{1mm} \centering
\includegraphics[width=9cm]{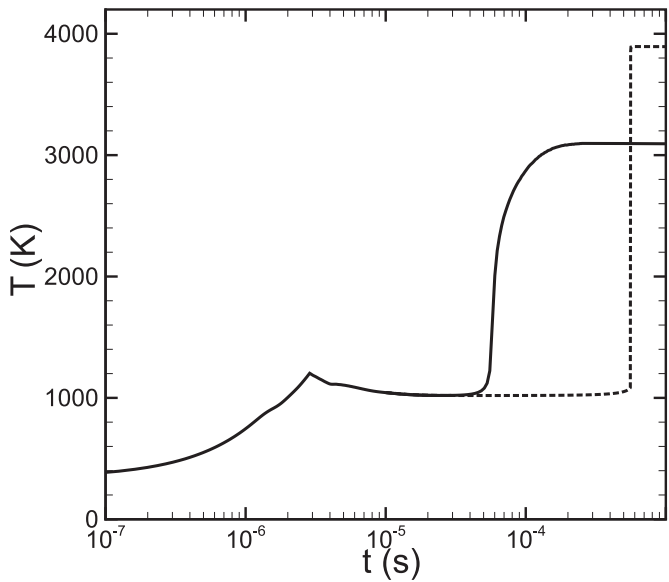}
\caption {\label{Fig5} Temperature evolution in
the center of the hot spot for case of low energy deposition
(${t_a} < \Delta {t_Q} < t_{ind}$). $L = 1$mm, $ \Delta {t_Q} = 16\mu $s
(solid line), ${P_0} = 10$atm  (dashed line).}
\end{figure}
\begin{figure}
\vspace*{1mm} \centering
\includegraphics[width=9cm]{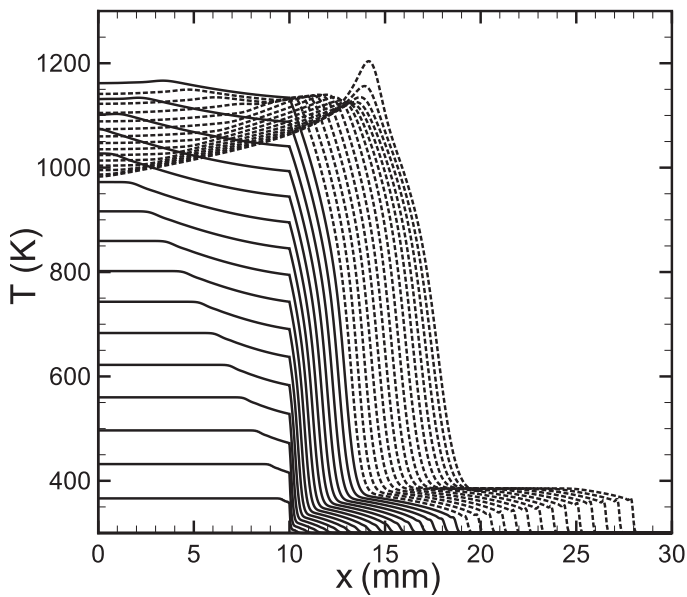}
\caption {\label{Fig6} Evolution of temperature profiles
in the hot spot during energy deposition (solid lines)
and during induction phase (dashed lines). $L = 1$cm,
$ \Delta {t_Q} = 16 \mu $s.
Profiles are presented for time instants
with interval $ \Delta {t} = 1 \mu $s .}
\end{figure}
\begin{figure}
\vspace*{1mm} \centering
\includegraphics[width=9cm]{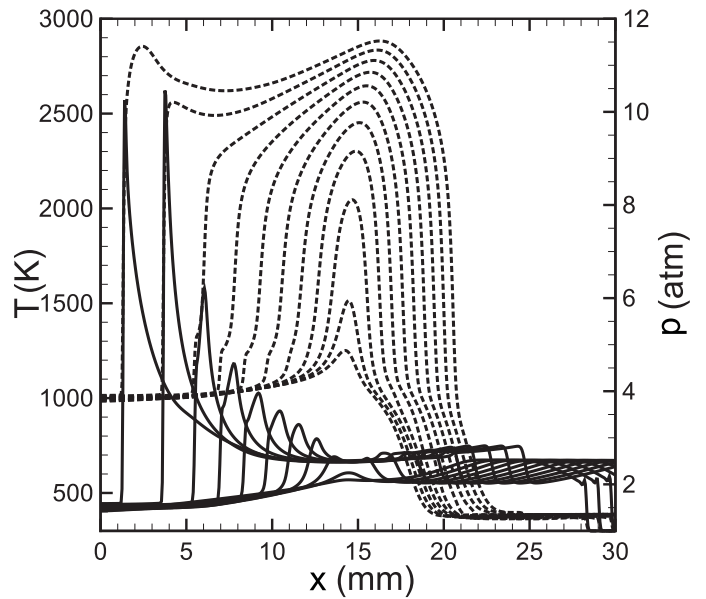}
\caption {\label{Fig7}
Evolution of temperature (dashed lines) and pressure
(solid lines) profiles illustrating ignition process
in the energy release region,  $L = 1$cm,
$ \Delta {t_Q} = 16 \mu $s.
Profiles are presented for time instants
with interval $ \Delta {t} = 1 \mu $s .}
\end{figure}

In case when the ignition time is greater
than acoustic time and the energy addition
time is less than acoustic time $ \Delta {t_Q} < {t_a} < {t_{ign}}$
the gradient induced by the rarefaction wave forms on the
stage after the end of energy addition.
During the relatively long induction phase the
acoustic perturbations equalize pressure in the
energy deposition zone and further ignition
processes evolve at constant pressure from
the steady temperature gradient.
Temperature in the top of gradient remains
nearly constant till the ignition takes place (see Fig.~\ref{Fig5}).
The combustion regime forming in this case depends on
the environmental conditions (e.g. pressure \cite {LKI6}):
at ${P_0} = 1$atm such a temperature gradient causes
deflagration wave formation, at ${P_0} = 10$atm it causes detonation wave.

An interesting scenario takes place when the energy
deposition time is slightly longer then acoustic
time, $ \Delta {t_Q} = 20\mu $s,
so that $ {t_a} < \Delta {t_Q} < {t_{ign}}$.
At the beginning the gasdynamics
process is similar to the previous case. However,
the heating time is longer, and the rarefaction
wave has time to reach the left wall, to be reflected
from the wall and return to the edge of the hot spot
during time of the energy deposition (§15 in Ref.~ \cite {LL16}).
As a result even before the reaction started
the hot spot expands with the temperature profile
in the mixture consisting of a shallow temperature
gradient in the direction to $x > 0$  and a steep temperature
gradient in the direction $X =0$. The initial stage of the
temperature time evolution, before the reaction has
started, is depicted in Fig.~\ref{Fig6}.
By the time the reactions begins at the top of
the gradients, the right side gradient is too
steep to trigger a detonation. Instead a
deflagration wave propagating to the right
is ignited Ref.~ \cite {LKI6}). At the same time due to the
elevated pressure the left shallow gradient
can facilitate triggering detonation through
the Zeldovich gradient mechanism. The corresponding
time evolution of temperature and pressure profiles
showing propagation of the spontaneous wave to the
left along the temperature gradient, its
coupling with the pressure waves and detonation
initiation as well as deflagration initiation by
the right steep temperature gradient
are depicted in Fig.~\ref{Fig7}.

The rapid energy deposition such that the
heating time, $ \Delta {t_Q}$, is comparable to
the characteristic acoustic time scale of the volume, ${t_a}$,
always results in the shock waves propagating away from the hot spot.
A particular scenario of the resulting combustion regime
depends on the size of the hot spot, though the basic
physics appears to be similar to that described above.
In the case of the smaller size of the hot spot ($L = 1$mm)
scenario of the combustion regime ignition may differ
because the scale of the temperature gradient created by
the rarefaction wave may not be compatible with the detonation
formation in real mixtures. There are many different
scenarios that include direct detonation formation by a
strong enough shock wave in the context of a thermal
explosion, or the shock waves propagating away from the
hot spot producing ignition of the fast deflagration
propagating behind the shock waves
(regime 3 according to the classification in Ref.~\cite {LKI6}).

\section{Combustion regimes: millisecond time scale of energy deposition}

Sufficiently rapid and large amount of thermal energy
deposition into a reactive gas can trigger either
direct initiation of detonation through a constant
volume explosion, or through the Zeldovich gradient mechanism
due to the shallow gradient formed by the rarefaction wave at
the increased pressures in the hot spot region.
In case of the rapid but relatively small energy deposition
the resulting regime is a fast deflagration wave Ref.~\cite {LKI6}.
The scenario for low power thermal energy addition
over a longer period of time is different.
If the acoustic time is much less than the energy deposition
time, ${t_a} \ll \Delta {t_Q} \leq {t_{ign}}$ ,
then there is enough time for pressure to be spatially
homogenized by acoustic waves. In this case there are
no strong compression waves emitted from the hot spot,
and the combustion regimes initiated by the
energy deposition into the hot spot depend
essentially on the steepness of the temperature gradient,
which is formed by the thermal wave and gas expansion
in the vicinity of the hot spot.

\begin{figure}[t]
\vspace*{1mm} 
\includegraphics[width=8cm]{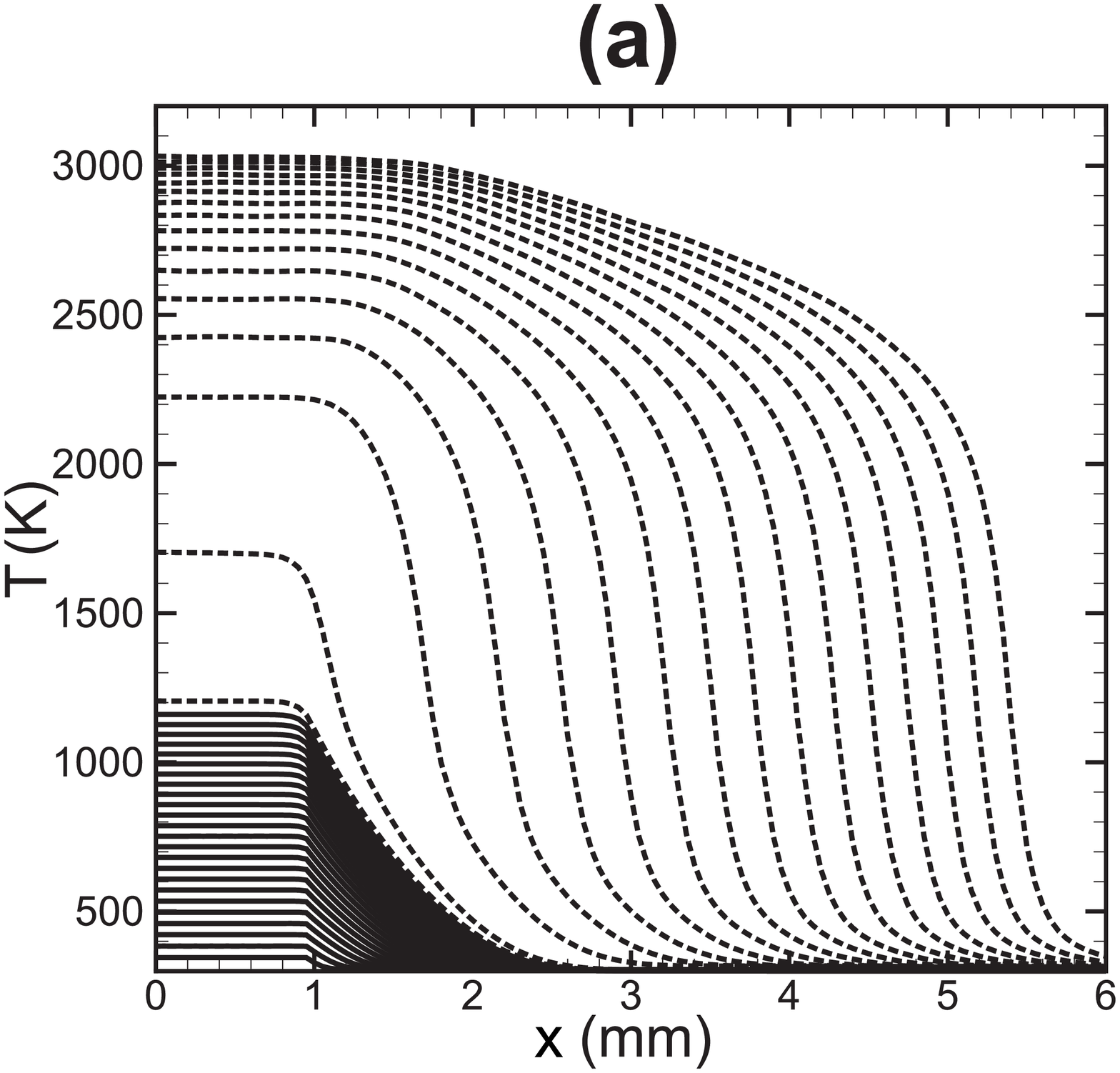}
\includegraphics[width=8cm]{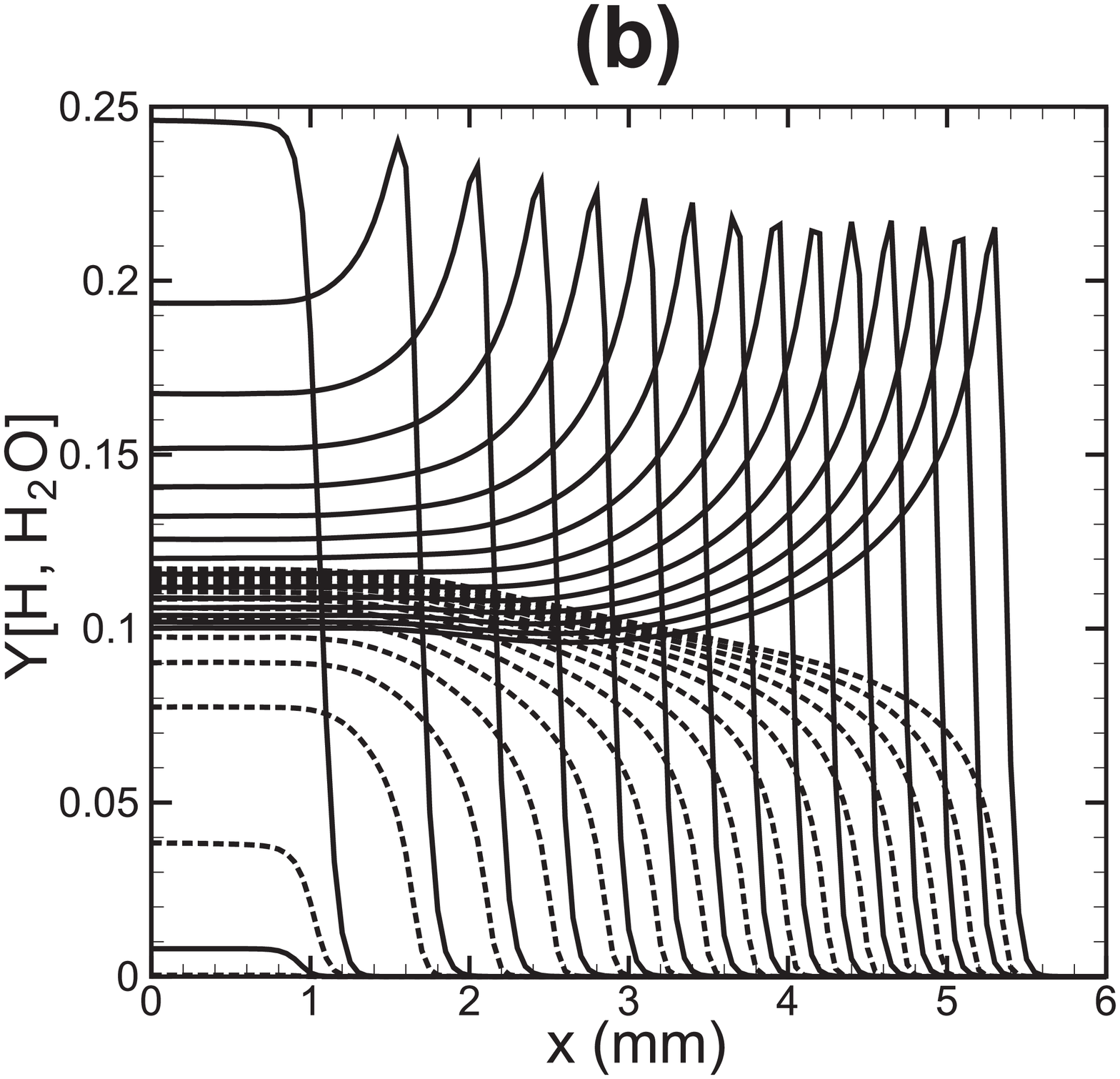}
\caption {\label{Fig8} (a) Evolution of temperature profiles in the hot spot during energy deposition (solid lines) and after it during the combustion wave formation (dashed lines). (b)Evolution of $H_2O$ concentration (dashed lines) and $H$-radical concentration (solid lines) profiles illustrating combustion wave formation on the gradient formed in the energy release region.$L = 1$mm, $ \Delta {t_Q} = 1000\mu $s. Profiles are presented for time instants with interval $ \Delta t = 5\mu$ s.}
\end{figure}

During the time of energy deposition $ \Delta {t_Q}$
the thermal wave propagates away from the hot spot at
the distance ${x_T}/mm = {(\chi \Delta {t_Q})^{1/2}}
 \approx 0.9{(\Delta {t_Q}/ms)^{1/2}}$ .
Some of the mass in the volume of the hot spot heated
by the added energy, therefore of higher temperature,
flows away as the temperature increases and the density falls.
The expelled mass together with the thermal wave give rise
to the temperature gradient in the surrounding mixture
behind the boundary of the hot spot. The temperature
profile is almost linear because of the weak temperature
dependence of the coefficient of thermal conduction
($ \kappa  \propto {T^{0.75}}$).
Fig.~\ref{Fig8}a shows the temperature gradient formed at the end of
the energy deposition for $ \Delta {t_Q} = 1000\mu $s, $ L = 1$mm
(solid lines).
A greater distance compared to that created by
the thermal wave alone is due to the hot spot
expansion, the decreased density and increased
temperature in the hot spot during the process of
energy deposition. However, according
to Ref.~\cite {LKI6} this type of linear
temperature gradient is too steep to initiate
detonation through the Zeldovich gradient
mechanism and as a result a deflagration
wave is initiated. The thermal wave and
the gas expansion are too slow to expand
temperature and to form a temperature gradient
compatible with the detonation formation in real
mixtures at atmospheric or lower pressures Ref.~\cite {LKI6}.
Long before the thermal wave moves away a sufficiently
long distance the temperature of the mixture rises
to ignite the reaction, so that either
a deflagration wave or a fast deflagration wave
are initiated according to the classification
of Ref.~\cite {LKI6}. In Fig.~\ref{Fig8}a dashed lines
show the deflagration wave formation out
from the formed temperature gradient. Fig.~\ref{Fig8}b represents the evolution of $H$-radical and $H_2O$ concentration profiles while the combustion wave is forming.

While speed of sound does not depend on pressure,
the induction time ${t_{ind}}(T)$  at the
temperature range $(1100 \div 1200)$K
is considerably longer at larger pressures (see Fig.~\ref{Fig1}).
This leads to a significant delay of ignition for
the same energy deposition regime at higher pressure.
As an example, the growth of temperature at the center of
the hot spot $L = 1$mm  and at instants when the reaction
starts are shown in Fig.~9 for
initial pressures ${P_0} = 1$atm  and ${P_0} = 10$atm and
for $\Delta {t_Q} = 1$ms.
\begin{figure}
\vspace*{1mm} \centering
\includegraphics[width=9cm]{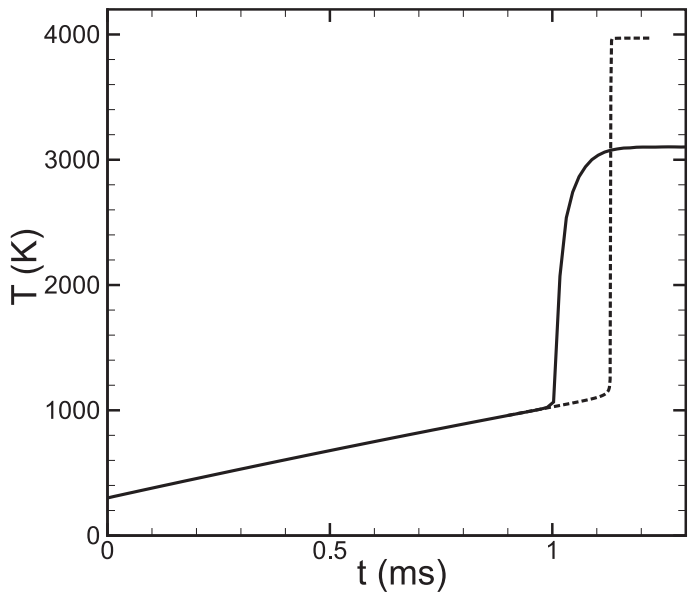}
\caption {\label{Fig9} Temperature evolution in the
center of the hot spot for case of slow
energy deposition (${t_a} \ll \Delta {t_Q}$). $L = 1$mm, ${P_0} = 1$atm
(solid line), ${P_0} = 10$atm  (dashed line). $\Delta t_Q=1000\mu$s for $P_0=1$atm and $1100\mu$s for $P_0=10atm$.}
\end{figure}
\begin{figure}[t]
\vspace*{1mm} 
\includegraphics[width=8cm]{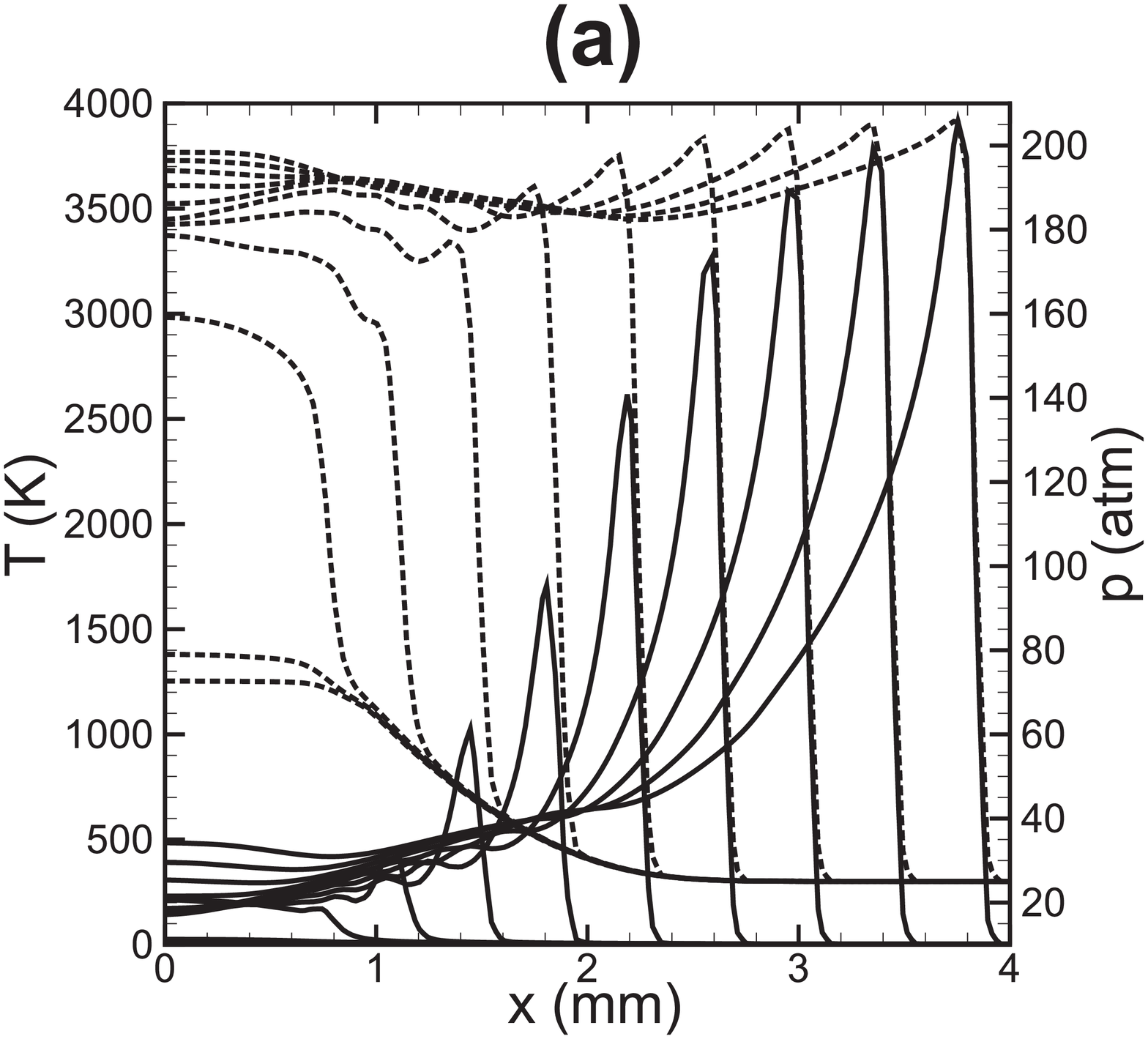}
\includegraphics[width=7.5cm]{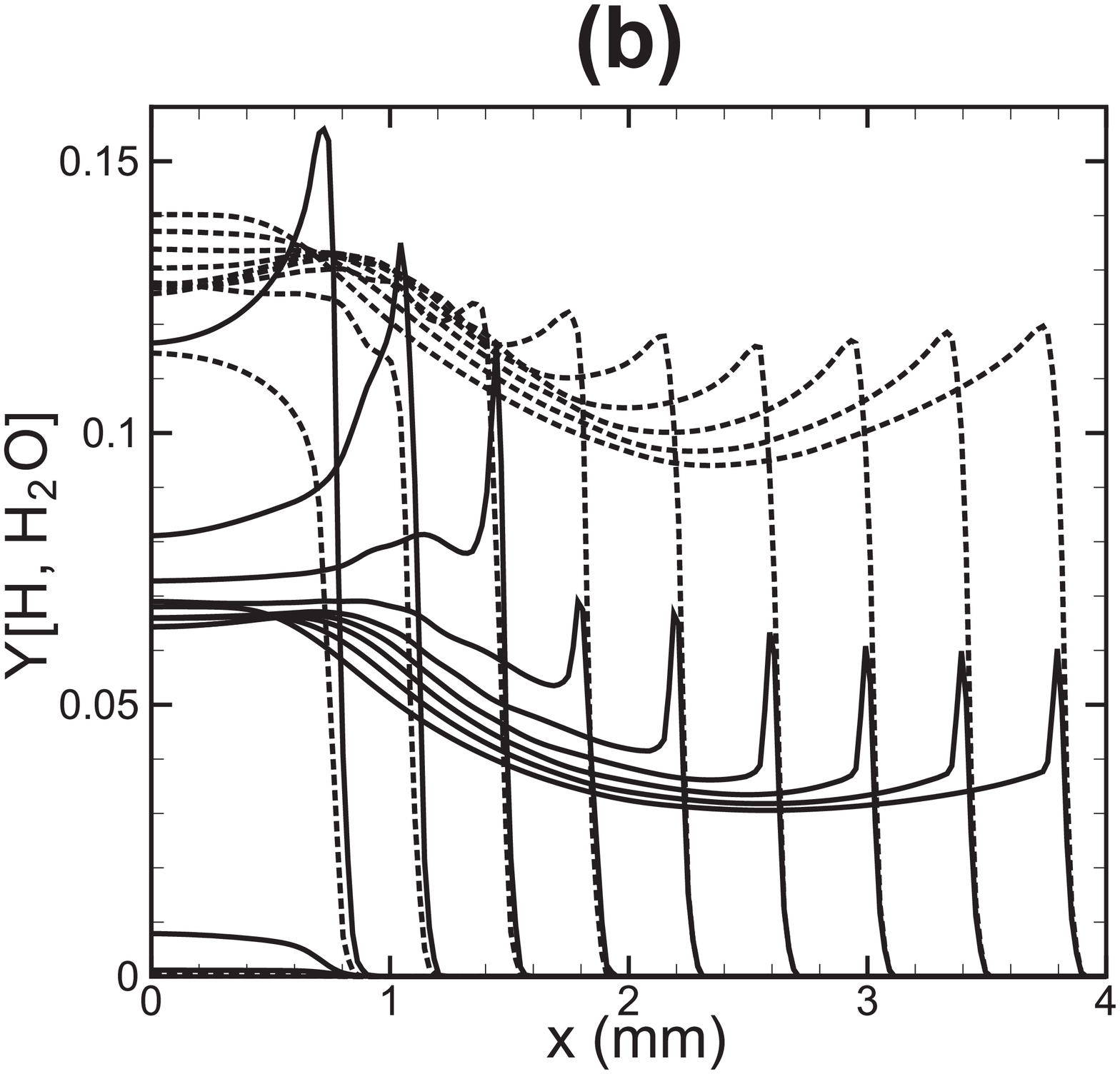}
\caption {\label{Fig10}
(a) Evolution of temperature (dashed lines) and
pressure (solid lines) profiles illustrating
detonation formation on the gradient formed in
the energy release region. (b) Evolution of $H_2O$ concentration (dashed lines) and $H$-radical concentration (solid lines) profiles illustrating detonation formation on the gradient formed in the energy release region. $L = 1$mm,
$\Delta {t_Q} = 1100\mu $s, ${P_0} = 10$atm.
Profiles are presented for time instants
with interval $\Delta t = 2\mu $s.}
\end{figure}

Since the coefficient of  thermal conductivity
does not depend on pressure, and the steepness of
the temperature gradient for direct detonation
initiation through the Zeldovich gradient mechanism
decreases considerably with the increase in
pressure Ref.~\cite {LKI6}, the temperature gradient
created by the thermal wave can trigger detonation at
high enough initial pressure. To elucidate the process
we consider a relatively small hot spot of size,  $L = 1$mm
at initial pressure ${P_0} = 10$atm.
For a sufficiently long energy deposition time,
the thermal wave creates gradient appropriate for
the detonation initiation through the Zeldovich mechanism
at ${P_0} = 10$atm Ref.~\cite {LKI6} (dashed line in Fig.~\ref{Fig9}).
Figure~\ref{Fig10}a shows the calculated temporal evolution of
temperature and pressure profiles illustrating formation of
the temperature gradient outside of the hot spot,  $L = 1$mm,
the development of the spontaneous wave along the gradient
and transition to detonation for the energy
deposition time $\Delta {t_Q} = 1$ms. Fig.~\ref{Fig10}b shows the corresponding evolution of the concentration profiles for $H$-radicals and the combustion products ($H_2O$).

\section{Energy of ignition}

The results obtained clearly allow us to estimate
the energy required for ignition of a particular
combustion regime. At the same time it should be
noted that the amount of the ignition energy
obtained by extrapolating results of the one-dimensional
problem most likely will not match the actual value of
the ignition energy for the three-dimensional problem
where the process is associated with a three-dimensional
expansion and converging rarefaction wave. This difference
is particularly important for the initiation of detonation.
In this case the three dimensional expansion additionally
enhances the rarefaction, leading to less suitable
conditions for a detonation  initiation.
A large number of 3D simulations of ignition due to the energy addition to the spherical hot-spot were used to verify realization of different combustion regimes, and to compare the ignition energy obtained from 3D spherically symmetric model to that extrapolated using the 1D model and to assess effect of spherical expansion on the ignition process.

\begin{figure}
\vspace*{1mm} \centering
\includegraphics[width=9cm]{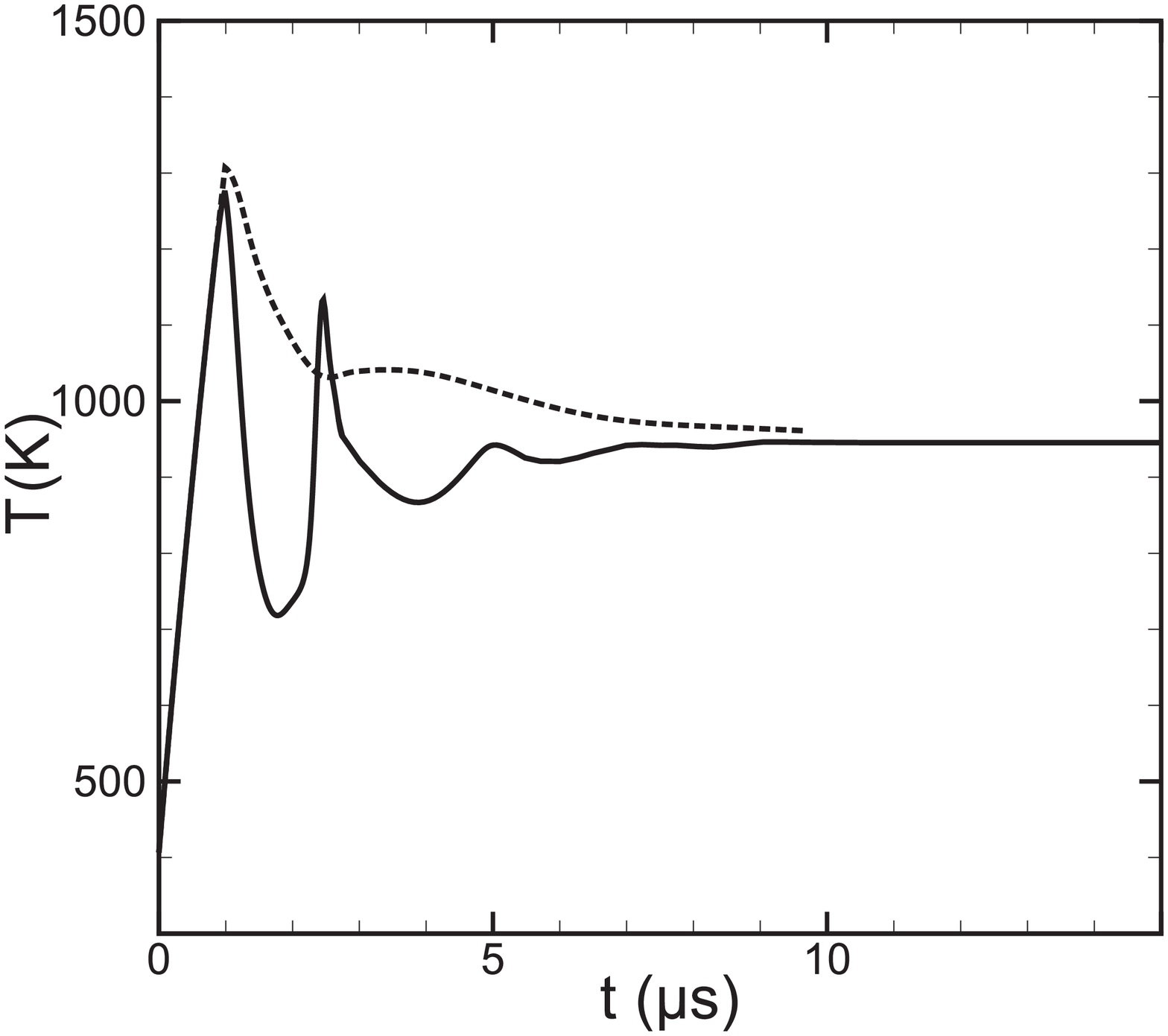}
\caption {\label{Fig11}
Temperature evolution in the center of
the hot spot for case of rapid energy deposition
(${t_a} > \Delta {t_Q}$) in one-dimensional case
(dashed line) and three-dimensional
case (solid line), $L = 1$mm.}
\end{figure}
\begin{figure}
\vspace*{1mm} \centering
\includegraphics[width=9cm]{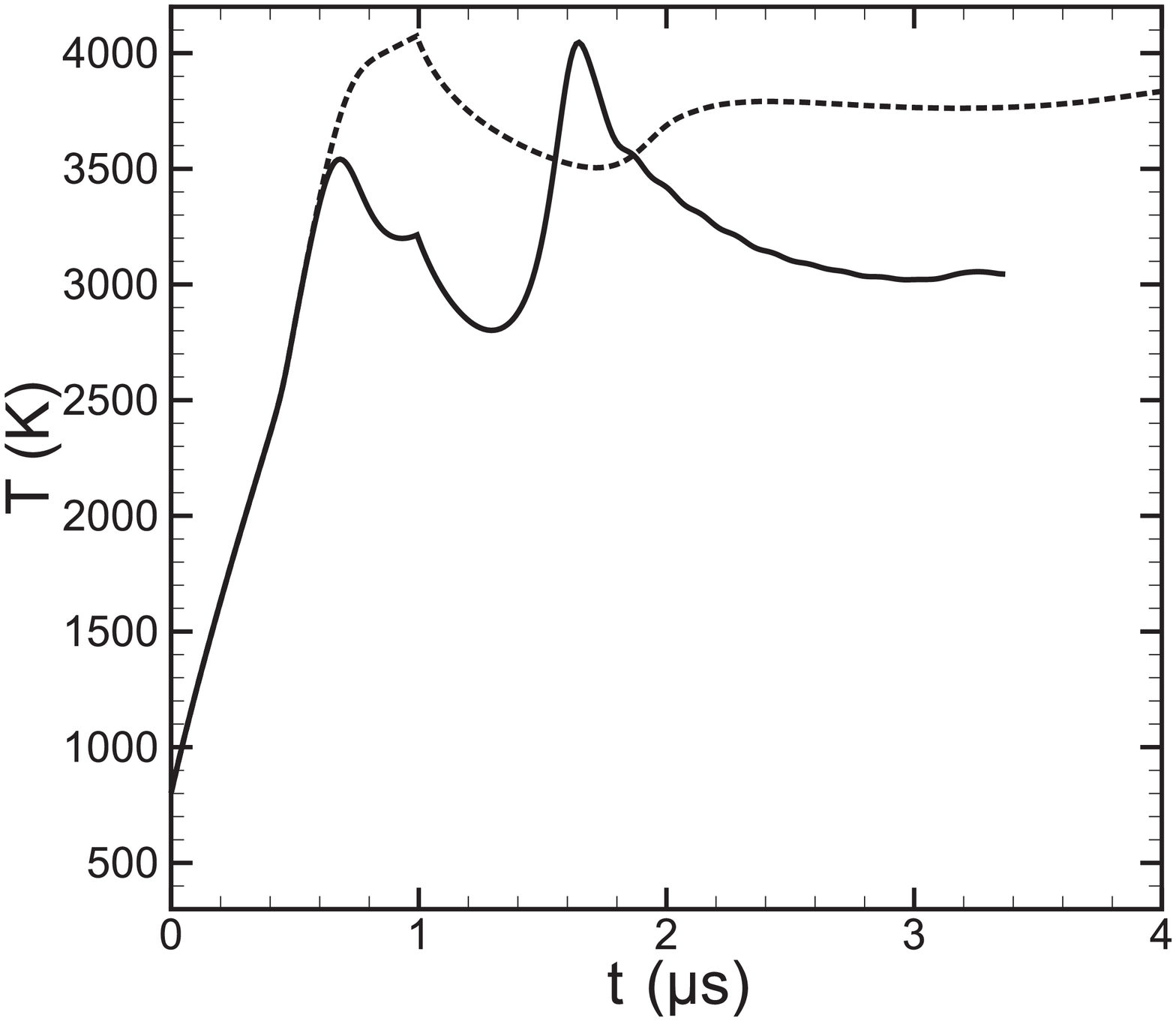}
\caption {\label{Fig12}
Temperature evolution in the center of
the hot spot for case of rapid energy
deposition ($t{}_a > \Delta {t_Q}$)
in one-dimensional case (dashed line) and
three-dimensional case (solid line); $L = 1$mm.}
\end{figure}

Figure~\ref{Fig11} shows the temperature evolution in the center of
the hot spot for rapid energy deposition (${t_a} \sim \Delta {t_Q}$)
in a one-dimensional case where $L = 1$mm  and in the three-dimensional
spherical hot spot of radius $R = 1$mm.
In both cases the final temperature in the center of
the hot spot tends to the same value.
Temperature oscillations in the spherical
hot spot are caused by  convergence of the rarefaction wave
to the center and reflection, which is enhanced by
the expansion of the gaseous spherical hot spot,
where the density decreases approximately
as $ \rho  \propto {\left( {1 - {{(r/R)}^2}} \right)^{\frac{1}{{\gamma  - 1}}}}$. Since the rarefaction is much stronger
for the 3D spherical expansion, the final
temperature of the spherical target is considerably
lower than in the planar case at the same parameters
of energy deposition if the rarefaction reflects
before the energy is deposited (${t_a} < \Delta {t_Q}$).
For the same reason, the actual energy required to initiate
a detonation is larger for the spherical target
than that in the planar case. Even if the temperature
of the hot spot rises high enough for initiating detonation
in the planar case, it is reduced by the converging and
reflecting rarefaction wave and detonation can not be ignited.
An example of such a scenario is shown in Fig.~\ref{Fig12},
where for the same conditions the detonation is
initiated in the plane geometry and
it is not for a spherical target.

\begin{figure}
\vspace*{1mm} \centering
\includegraphics[width=9cm]{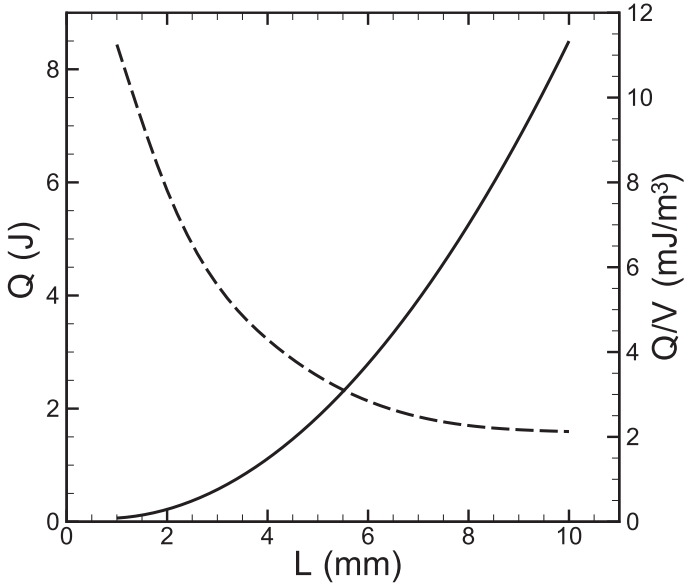}
\caption {\label{Fig13}
The minimal energy required for detonation
initiation in H2-O2 depending on the size
(radius) of the hot spot for $\Delta {t_Q} = 1 \mu $s.
Solid line is total deposited energy;
dashed line is the specific energy
($Q/V{\rm{[mJ}} \cdot {\rm{m}^3}$).}
\end{figure}

The larger the acoustic time ${t_a}$
compared to the time of energy deposition
the less the influence of the rarefaction wave
on the detonation initiation.
This means, for example, that with the increase of
the hot spot size (note, that ${t_a} \propto L$)
the initiation of detonation requires less energy
deposition into the specific volume of the hot spot
for a given time of power deposition $\Delta {t_Q}$,
although the total deposited energy will be larger
than in the case of detonation initiation in a
smaller hot spot using a higher level of power deposition.
This tendency is illustrated in Fig.~\ref{Fig13}.
For a shorter period of power deposition the energy amount per unit of volume
capable for detonation initiation decreases.
The lower limit for hydrogen detonation initiation can be
obtained using a short sub-microsecond laser pulse focused
in a sub-millimeter area (see e.g. Ref.~\cite {Lewis27})
and it is estimated as $ \sim {10^{ - 2}}$mJ,
which agrees with the extrapolation of the dependence shown in Fig.~\ref{Fig13}.

\begin{figure}
\vspace*{1mm} \centering
\includegraphics[width=9cm]{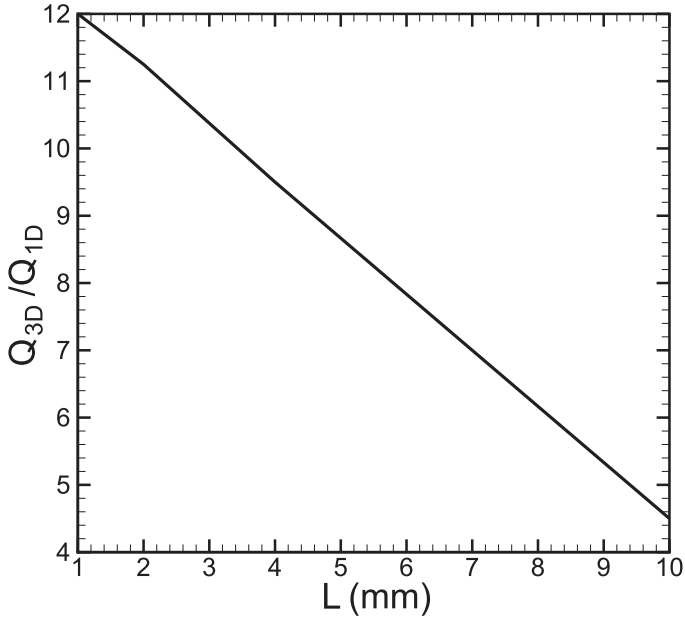}
\caption {\label{Fig14}
The ratio of energies required for detonation
initiation in ${H_2}-{O_2}$ obtained for 3D and
1D model versus the size (radius) of
the hot spot for $\Delta {t_Q} = 1\mu$.}
\end{figure}

\section{Discussion and Conclusions}

In the present paper we used detailed chemical
kinetics and transport models to study consequences
of localized transient energy deposition into
a stoichiometric mixture of hydrogen-oxygen
leading to the ignition of different regimes of
combustion. It is shown that depending on the
parameters of energy deposition
(deposited energy amount, deposition time
scale and size of the hot spot) there are
two main mechanisms of reaction wave initiation:
the Zeldovich gradient mechanism Ref.~\cite {Zeld19802}
 and  the volumetric thermal explosion (which actually
represents one of the asymptotics of
the Zeldovich mechanism for the gradient of zero steepness).
For practically important time scales
the principal scenarios of ignition are:
1) for sub-microsecond pulses the volumetrical
explosion takes place inside the hot spot;
2) for microsecond pulses the gradient of
temperature and pressure arises on the profile
created by the rarefaction wave and ignition
starts via Zeldovich mechanism on the gradient of induction time;
3) for millisecond pulses gasdynamical expansion gives
rise to a temperature gradient at approximately constant
pressure, and the ignition starts according to
the Zeldovich mechanism Ref.~\cite {Zeld19802} with all the features
inherent to chain-branching chemistry disclosed in Ref.~\cite {LKI6}.
In the three-dimensional case spherical expansion of
the hot spot weakens the generated shock wave in favor
of an intensified rarefaction wave. It results in
sufficient drop in temperature and pressure in the
hot spot on the time scales of the order of acoustic time.
Thus, for the same conditions as in one-dimensional case,
a less intensive combustion regime arises.
The deflagration regimes are less sensitive.
However to obtain detonation one should increase
sufficiently the power of the energy source.
For example for $L = 1$mm  and $\Delta {t_Q} = 1\mu$s
the energy amount
is about 10-12 times larger compared to
that obtained via extrapolating of the results of 1D model.
This is clearly seen from Fig.~\ref{Fig14}, which shows the
ratio of energy required for the detonation
initiation in 1D and 3D cases. The calculations were done for the fixed time of the energy deposition $\Delta {t_Q} = 1\mu$s, for the planar and
spherical hot spots of different sizes
( $L$ in 1D case and $R + L$  for 3D case). In particular,
it is seen that with the increase of
the hot spot size, and corresponding increase of
the acoustic time ${t_a}$,
the role of a rarefaction wave becomes less important.

\begin{acknowledgements}
This research was supported by the Grant of Russian
Ministry of Science and Education "Non-stationary combustion regimes
for highly efficient and safety energy production"
(Program 1.5/XX, Contract No. 8648) and
partially supported by the EC FP7 project ERC PBL-PMES (No. 227915).
The authors appreciate valuable discussions with V. E. Fortov.
\end{acknowledgements}

\end{document}